\documentclass[useAMS,usenatbib]{mn2e}

\usepackage{graphicx}
\usepackage{amsmath}

\title[Practical Cosmology with Lenses]{Practical Cosmology with Lenses}
\author[Stephen Eales]{S.A. Eales$^{1}$\thanks{E-mail:
sae@astro.cf.ac.uk}\\ 
$^{1}$School of Physics and Astronomy, Cardiff University,\\ 
The Parade, Cardiff CF24 3AA, UK}
\begin{document}

\date{Accepted by MNRAS}

\pagerange{\pageref{firstpage}--\pageref{lastpage}} \pubyear{2002}

\maketitle

\label{firstpage}

\begin{abstract}
Surveys with submillimetre telescopes are revealing
large numbers of gravitationally lensed high-redshift sources. I
describe how, in practice, 
these lensed systems could be simultaneously used to estimate the values
of cosmological parameters, 
test models for the evolution of the distribution of dark-matter halos 
and investigate the properties of the source population. 
Even 
the existing sample of lenses found with the Hershcel Space Observatory is 
enough to formally rule out the standard models
of the evolving population of dark-matter halos, 
with the likely explanation a combination of baryon
physics and the perturbation by infalling baryons of the density distribution of dark matter
at the centres of the halos.
Independently of the evolution of the halos, observations 
of a sample of 100 lensed systems would be enough to 
estimate $\Omega_{\Lambda}$ with a precision of 5\% and 
observations of 1000 lenses would be enough to estimate $w$, 
the parameter in the equation-of-state of dark energy,
with a precision similar to that obtained from the Planck 
observations of the cosmic microwave background. 
While the fraction of submillimetre sources that are lensed 
depends weakly on the 
specific halo mass function that is used in the model, it depends very strongly on the 
evolution of the submillimetre luminosity function of the source population. Therefore
measurements of the lensing fraction
could be used to investigate galaxy evolution in a way that is independent
of the properties of the intervening halos.

\end{abstract}


\section{Introduction}

It has been realised for many years that 
large samples of gravitational lenses have huge potential power for cosmological 
investigations, including measurements of fundamental cosmological parameters 
(e.g. Kochanek 1992, 1996; Grillo et al. 2008; Oguri et al. 2012), investigations of 
the evolution of the equation-of-state 
of dark energy (Zhang, Cheng and Wu 2009) and tests of 
theories of modified gravity (Zhao, Li and Koyama 2011). Also for many years, 
it has been predicted that submilletre surveys would be the way of assembling these 
large samples (Blain 1996; Perotta et al. 2002, 2003; Negrello et al.
2007). Recently observations with 
the Herschel Space Observatory (Negrello et al. 2010;
Wardlow et al. 2013) and the South Pole Telescope (Vieira et al. 2013) 
have confirmed these predictions, showing that samples of hundreds and 
even thousands of lensed systems (Gonzalez-Nuevo et al. 2012) are within reach.

Apart from the potential number of lensed sources, there are a number of other advantages
for cosmology of lensed sources found using this technique. First, the sources are 
generally at much higher redshifts than those found using other techniques, making them 
more sensitive for cosmological experiments
(Bussmann et al. 2013; Weiss et al. 2013). Second, the 
sources, but not the lenses, are bright at submillimetre and radio wavelengths, 
making it easy to map them and to distinguish the lens and the source emission. Third, 
galaxies are optically-thin at radio and submillimetre wavelengths, meaning that no 
correction has to be made for absorption by the lens, a problem with optical techniques.

Given this huge potential, it seems the right moment to take a severely practical look 
at what one can actually measure. The properties of a sample of lensed sources actually 
depend on four things: the cosmological model, the population of sources, the statistical 
distribution of dark-matter halos expressed as a function of mass and redshift, 
and the density distributions of individual halos. A fifth can also be added 
if one allows for the possibility of modifying gravity. An important question to 
address is the best way to use the samples of lenses to investigate separately 
the different things on which their properties depend. There is also a 
second question that is vital for any practical cosmological investigation. What 
are the possible systematic effects?

In this paper, I address these questions. All models for the 
evolution of galaxies start with a statistical distribution of dark matter halos as 
a function of mass and redshift. This distribution is thus vital for our 
understanding of the evolution of galaxies. It has either been inferred from 
analytic arguments (Press and Schechter 1974) or by fitting analytic 
functions to the results of N-body simulations of the evolution of dark 
matter (Sheth and Tormen 1999; Tinker et al. 2008), but there has been no 
way to test these predictions observationally apart from the very indirect and 
unsatisfactory method of comparing the results of the galaxy evolution models which 
are based on these functions to observations. In Section 2 I show 
that if we assume the standard cosmological model, which has been measured with great 
precision by Planck and many other cosmological experiments (Planck Collaboration 2013), 
it is possible to test these predictions irrespective of the properties of 
the source population. I test the predictions of the models against the observations
of a large sample of Herschel lensed sources (Bussmann et al. 2013), showing that the
observations will have ample statistical power to set useful constraints on the
halo mass function or alternatively on the physics of the infalling baryons. 

Grillo et al. (2008) proposed a method for estimating cosmological parameters 
that would be independent of the distribution of dark-matter halos and of 
the source population. In Section 3, I use realistic estimates of observational
precision
to estimate the accuracy with which one could measure 
$\Omega_{\Lambda}$ and $w$,  the parameter in the equation-of-state of dark energy.

I then consider the population of sources. In Section 4, I show 
there is one property of the source
population that depends exceptionally weakly on the
properties of the intervening population of lenses.
This is the fraction of submillimetre sources that are strongly lensed,
which in this paper is defined as a source that has a magnification factor $>2$.
This is, in principle, easy to measure since strongly lensed sources should have at
least two images, and 
I show that this property of submm samples 
is a sensitive measure of galaxy evolution.

In all the models I make two assumptions which simplify this initial analysis.
First, I assume that
the substructure within a lens does not have a significant effect on the lensing
properties. Although there is clearly substructure within dark-matter halos
(Diemand et al. 2008; Springel et al. 2008), the assumption that the substructure does
not have a large effect on the lensing properties is a reasonable one since the residuals found 
after fitting simple models to the data for individual lenses are invariably small
(Dye et al. 2014; Hezaveh et al. 2013b).
Lapi et al. (2012) also provide some theoretical support for this assumption, finding
that the difference between the overall halo mass function and one
that also contains subhalos differs by less than 5\% over the range of halo
mass $11.4 \leq log_{10}(M_H/M_{\odot}) \leq 13.5$.

The second assumption is that 
the density distributions of
all the lenses are represented by that of a singular isothermal sphere (SIS): 

\smallskip
\begin{align}
\rho(r) = {\sigma_v^2 \over 2 \pi G r^2}, 
\end{align}
\smallskip

\noindent where $\sigma_v$ is the line-of-sight velocity dispersion. There is a lot of
observational evidence 
that the density profiles of lenses, on the physical
scale of the lensing phenomena, are better represented by
this function than the NFW profile (Navarro, Frenk and White 1997) expected for dark-matter
halos (e.g Kochanek 1995; Koopmans et al. 2009; Bolton et al. 2012; Treu et al. 2010).
Furthermore, Lapi et al. (2012) have shown that a combination of the NFW profile
for dark matter and a stellar component with a Sersic profile actually gives
a density profile on the physical scale of the lensing very similar to that of equation (1). 
There is as yet relatively little information about the density profiles of the lenses
found in submillimetre samples, but Dye et al. (2014) have shown that the density
profiles of five Herschel lenses are similar to that of equation (1).

One assumption that turns out not to be important is about the sizes of the sources.
The effect of the sources not being point sources but having finite sizes is to
set an upper limit on the magnification factor. Perotta et al.
(2002) calculated that for sources with sizes in the range 1-10 kpc in 
the redshift range $z=1-4$ the maximum magnification lies in the range
10-30. I have 
taken into account the uncertainty in the sizes by investigating the effect on the
model predictions of changing the
maximum magnification, finding in practice that it has little effect
(Sections 2 and 4).

Unless otherwise stated, I assume the cosmological parameters from the 
Planck 2013 cosmological analysis (Planck Collaboration 2013): a spatially-flat universe
with $\Omega_M = 0.315$ and a Hubble constant of 67.3 $\rm km\ s^{-1}\ Mpc^{-1}$.

\section[]{Basic Lensing Formulae and the Distribution of Dark-Matter Halos}

In this section I assume that the values of the cosmological parameters
given by the latest analysis of the Planck team are correct (Planck Collaboration 2013).
This is a reasonable assumption to make, since our knowledge of the cosmological
parameters is much better than the observational constraints on the
evolving distribution of dark-matter halos, which are virtually non-existent. 
The
properties of the lensed sources also depend, of course, on the
properties of the source population. Short et al. (2012) suggest that
one prediction of a lensing model that is independent of the properties of the
source population is $P(z_l | z_s)$, the conditional probability of the redshift of the
lens given the redshift of the source. In this section I compare this prediction with
observations for three different distributions of dark-matter halos. One other property of
a lensed source that is relatively easy to measure (Bussman et al. 2013) is its
Einstein radius. Therefore I also use a second prediction of a lensing model that
is also independent of the properties of the source population: $P(\theta_E|z_s)$, the
conditional probability of the Einstein radius given the source redshift.

\begin{figure}
\includegraphics[width=84mm]{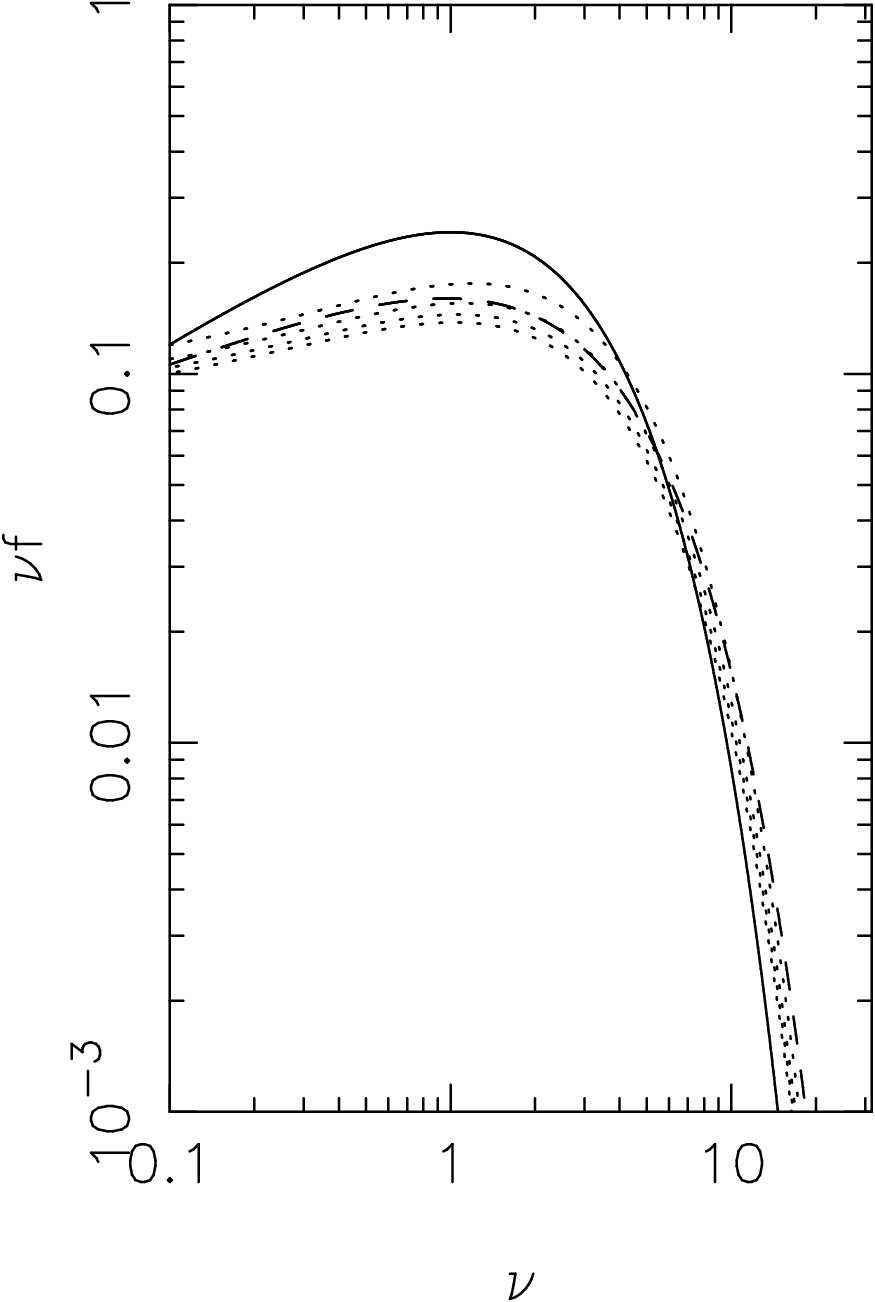}
  \caption{Plot of the analytic functions that we use for calculating the
halo mass functions: solid line---equation 8 (Press and Schechter 1974);
dashed line---equation 9 (Sheth and Tormen 1999); dotted lines---equation
10 (Tinker et al. 2008). The four lines shown for the halo mass function
from Tinker et al. (2008) are for redshifts of 0, 1, 2 and 3, with the function
decreasing in value with increasing redshift.} 
\end{figure}

\begin{figure}
\includegraphics[width=84mm]{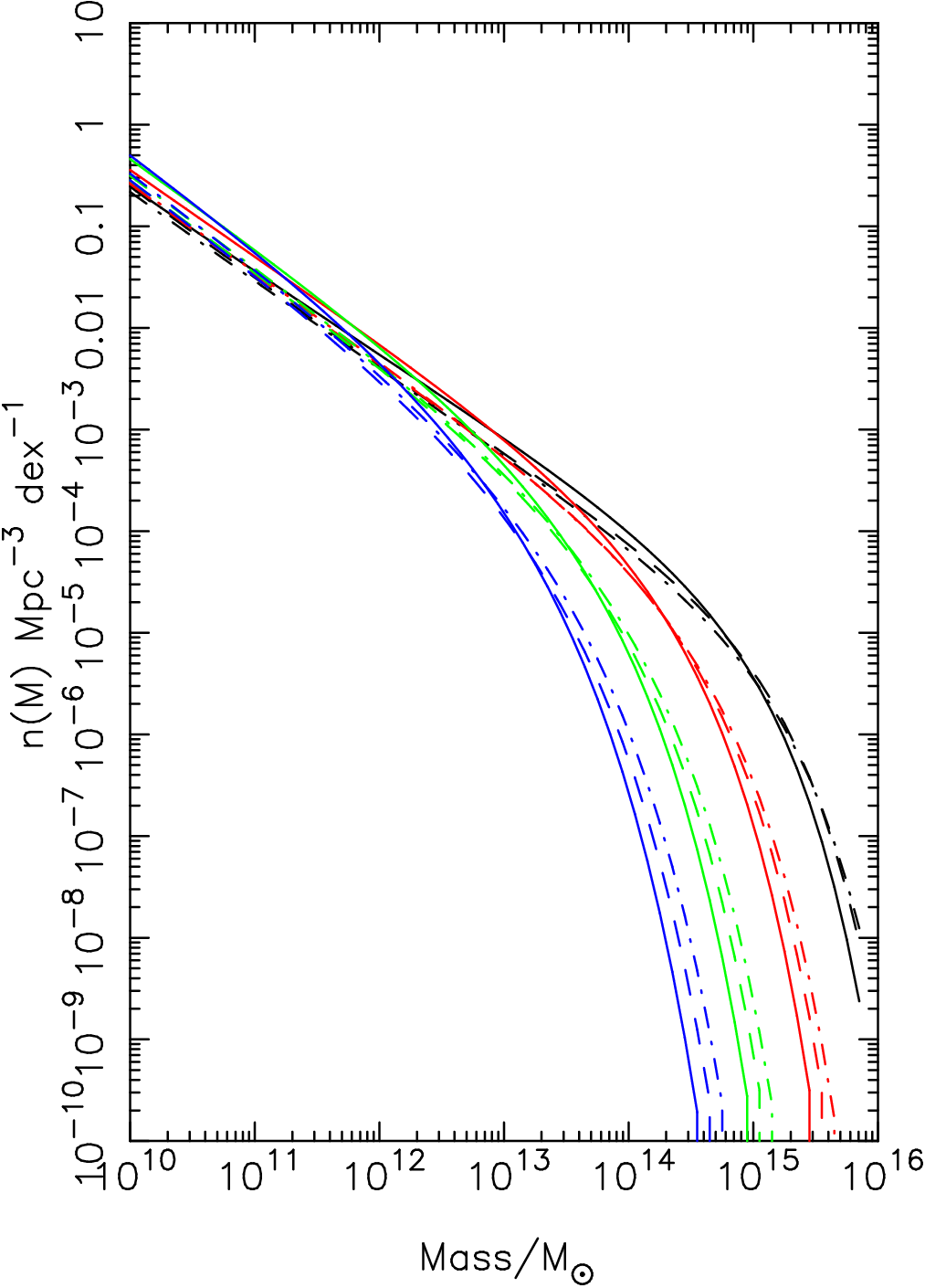}
  \caption{Plot of the halo mass function at four redshifts:
$z=0$ (black), $z=1$ (red), $z=2$ (green), $z=3$ (blue).
The solid lines show the halo mass function from
Press and Schechter (1974), the dashed lines show the
halo mass function from Tinker et al. (2008), and
the dot-dash line the halo mass function from
Sheth and Tormen (1999).}
\end{figure}

For a source that is being strongly lensed, defined as a source
with magnification $>2$, by a lens with an SIS
density profile, the probability of the magnification factor
$\mu$ is given by:

\smallskip
\begin{align}
p(\mu) = {8 \over \mu^3} 
\end{align}
\smallskip
\noindent (e.g. Peacock 1999). The probability that a source at redshift $z_s$ is lensed with a magnification
$\mu$ by a lens of mass $M$ at a redshift $z_l$ is given by:

\smallskip
\begin{align}
p(z_l, \mu, M | z_s) dz dM d\mu = &  \nonumber\\
  {c n(z_l,M)(1+z_l)^2 \over
  H_0 \sqrt{(1+z_l)^3 \Omega_M + \Omega_{\Lambda} } } & {d\sigma(>\mu) \over d\mu} dz dM d\mu  , 
\end{align}
\smallskip

\noindent in which $n(z,M)$ is the comoving number-density of dark-matter halos
as a function of mass and redshift and $\sigma(>\mu)$ is the cross-section
of a halo to gravitational lensing with magnification factor$>\mu$.
In this equation, the dependence on the mass of the halo comes in two
ways. First, the number-density of halos depends on mass. Second, the derivative
of the cross-section also depends on mass, as we will now see. For strong lensing
and an SIS profile, the derivative
of the cross-section is given by:

\smallskip
\begin{align}
{d\sigma(>\mu) \over d \mu} = {-8 \pi D_L^2 \theta_E^2 \over \mu^3}, 
\end{align}
\smallskip

\noindent where $\theta_E$ is the Einstein radius of the lens and $D_L$ is the
angular-diameter distance of the lens. The derivative of the cross-section depends
on mass because the Einstein radius depends on the mass of the halo.
This equation can be rewritten
as:

\smallskip
\begin{align}
{d\sigma(>\mu) \over d \mu} = {-8 \pi D_L^2 D_{LS}^2 \over \mu^3 D_S^2} \left({4 \pi \sigma_v^2 \over c^2}\right)^2, 
\end{align}

\smallskip

\noindent in which $D_{LS}$ and $D_S$ are the lens-source and source
angular-diameter distances (e.g. Peacock 1999). In this equation the derivative of the cross-section
is connected to the mass of the halo through the velocity dispersion.

An analytic expression commonly used to describe the
number-density of halos $n(M,z)$, originally derived by Press and Schechter
(1974), is:

\smallskip
\begin{align}
{M^2 n(M,z) \over <\rho> } {dM \over M} = \nu f(\nu) {d\nu \over \nu} 
\end{align}
\smallskip

\noindent In this expression $<\rho> = \Omega_M \rho_c$, the average
comoving density of the Universe, and the parameter $\nu$ is given
by

\smallskip
\begin{align}
\nu(M,z) = {\delta_c^2 / \sigma^2(M,z)}, 
\end{align}
\smallskip

\noindent in which $\delta_c$ is the critical ratio of density to average density
required for spherical collapse and $\sigma^2$ is the variance in the linear density
fluctuation field. I have calculated $\sigma^2$ using the transfer
function of Bardeen et al. (1986), normalising the field so that the value of
$\sigma^2$ at the current epoch 
in spheres of radius 8$h^{-1}$ Mpc is unity.

I have used three expressions for $f(\nu)$. The first comes from the
original back-of-the-envelope theoretical argument of Press and
Schechter (1974):

\smallskip

\begin{align}
f_{PS} (\nu) = \sqrt{ {1 \over 2 \pi \nu} } exp\left( {-\nu \over 2} \right) 
\end{align}

\smallskip 

\noindent The second is a modification to the Press-Schechter formula proposed
by Sheth and Tormen (1999) to give better agreement with the results
of N-body simulations of the formation of dark-matter halos:

\smallskip
\begin{align}
f_{ST} = {A \over \nu \sqrt{\pi} } (1 + {1 \over \nu'^p }) ({\nu' \over 2})^{1/2} exp( -\nu'/2) 
\end{align}
\smallskip

\noindent in which $\nu' = a \nu$, $a=0.707$, $p=0.3$ and $A=0.322$. The third function is also a fit
to the results of N-body simulations but one that allows for a change in the function with redshift
(Tinker et al. 2008):

\smallskip
\begin{align}
f_T(\nu) = {A \over 2 \nu } \left[ \left( {\delta_c \over b \sqrt{\nu} } \right)^{-\alpha}
+ 1 \right] exp( {\nu c \over \delta_c^2} ) 
\end{align}
\smallskip

\noindent where $A = 0.186(1+z)^{-0.14}$, $a = 1.47(1+z)^{-0.06}$, $b=2.57(1+z)^{-\alpha}$,
$\alpha=0.0107$ and $c=1.19$. Tinker et al. derived halo mass functions from N-body simulations
by finding halos with average densities a factor of $\Delta$ above the background density.
The halo mass function given here is for $\Delta=200$.
Note that I believe Short et al. (2012)
have used the wrong form for the halo mass function
of Tinker et al., which would explain the
large differences in their predictions for different
halo mass functions. Note the difference between their 
equation A5 and equation 10 in this paper.

Figure 1 shows these three functions, with the third function, the only one with a redshift
dependence, shown at four different redshifts. The three functions are quite similar. The
dependence on redshift of the third function is not strong, with a decrease of approximately
20\% in $f$ from a redshift of 0 to a redshift of 1. Figure 2 shows the number-density of
halos, $n(M,z)$, at four different redshifts for the three halo mass functions.
The strong dependence on redshift for all three halo mass functions is due to the
redshift dependence of the linear density field.

\subsection{A model connecting the masses and velocity dispersions of the halos}

The biggest problem in using the results of strong-lensing experiments to
test predictions for the halo mass function is in making the connection between
the small spatial scales on which the lensing occurs to the large scales on
which the halos are defined. Strong-lensing observations provide a measurement of the
mass interior to the Einstein radius, which for the lenses considered in this paper
is roughly equivalent to a projected radius of $\sim$10 kpc, whereas the halos
in N-body simulations
are selected on a scale of $\sim200$ kpc (Tinker et al. 2008). The lensing observations
provide a measurement of the mass in a core through the cluster, but a large extrapolation
is necessary to estimate the total mass of the halo.
We have used two different techniques for predicting the lensing signal from
the halo masses.

In this section I follow a number of authors (e.g. Mitchell et al.
2005; Short et al. 2012) in connecting the lensing signal to the masses
of the halos through the velocity dispersion of the halos.
Bryan and Norman (1998) 
derived a simple theoretical model connecting the one-dimensional
velocity dispersion of a halo at any redshift to its mass and then calibrated the
constant of proportionality in this relationship using N-body simulations.
This model also matches well the observed velocity-dispersion function of E/S0
galaxies at low redshift (Cirasuolo et al. 2005 and references therein), the morphological class into which most lenses
fall. I have used this model to derive the one-dimensional velocity dispersion for each halo, and
then use equations 3 and 5 to derive the lensing probability.
The relationship between the one-dimensional velocity dispersion and the halo mass derived
by Bryan and Norman (1998) is:

\smallskip
\begin{align}
\sigma_v(M,z_L) = 92.3[\Delta_{vir}(z_L)^{0.5} E(z_L) {M \over 10^{13} h^{-1} M_{\odot}} ]^{1/3} km/s, 
\end{align}
\smallskip

\noindent in which $E(z) = \sqrt{\Omega_M (1+z)^3 + \Omega_{\Lambda}}$, $\Delta_{vir}(z)
= 18 \pi^2 + 60[\Omega(z) - 1] - 32[\Omega(z) - 1]^2$ and
$\Omega(z) = \Omega_M (1 + z)^3 / E(z)^2$.

Figure 3 shows the probability of strong lensing ($\mu>2$) as a function of lens redshift for three
different source redshifts.
I have made the assumption that the maximum magnification is 50,
but the value chosen makes very little difference. In the SIS model, the steep decrease in 
the lensing probability with increasing magnification (equation 2)
means that the total lensing probability falls by only 4\% as the maximum magnification is reduced
from infinity to 10.
Figure 4 shows the probability of lensing as a function of halo mass for the different halo mass functions
and the same source redshifts. This figure shows that the halos that produce most of the lensing effect
have masses close to the knee of the halo mass functions, $\rm \simeq10^{13.5}\ M_{\odot}$, and therefore
similar to the halos of galaxy groups rather than individual galaxies. 

\begin{figure}
\includegraphics[width=84mm]{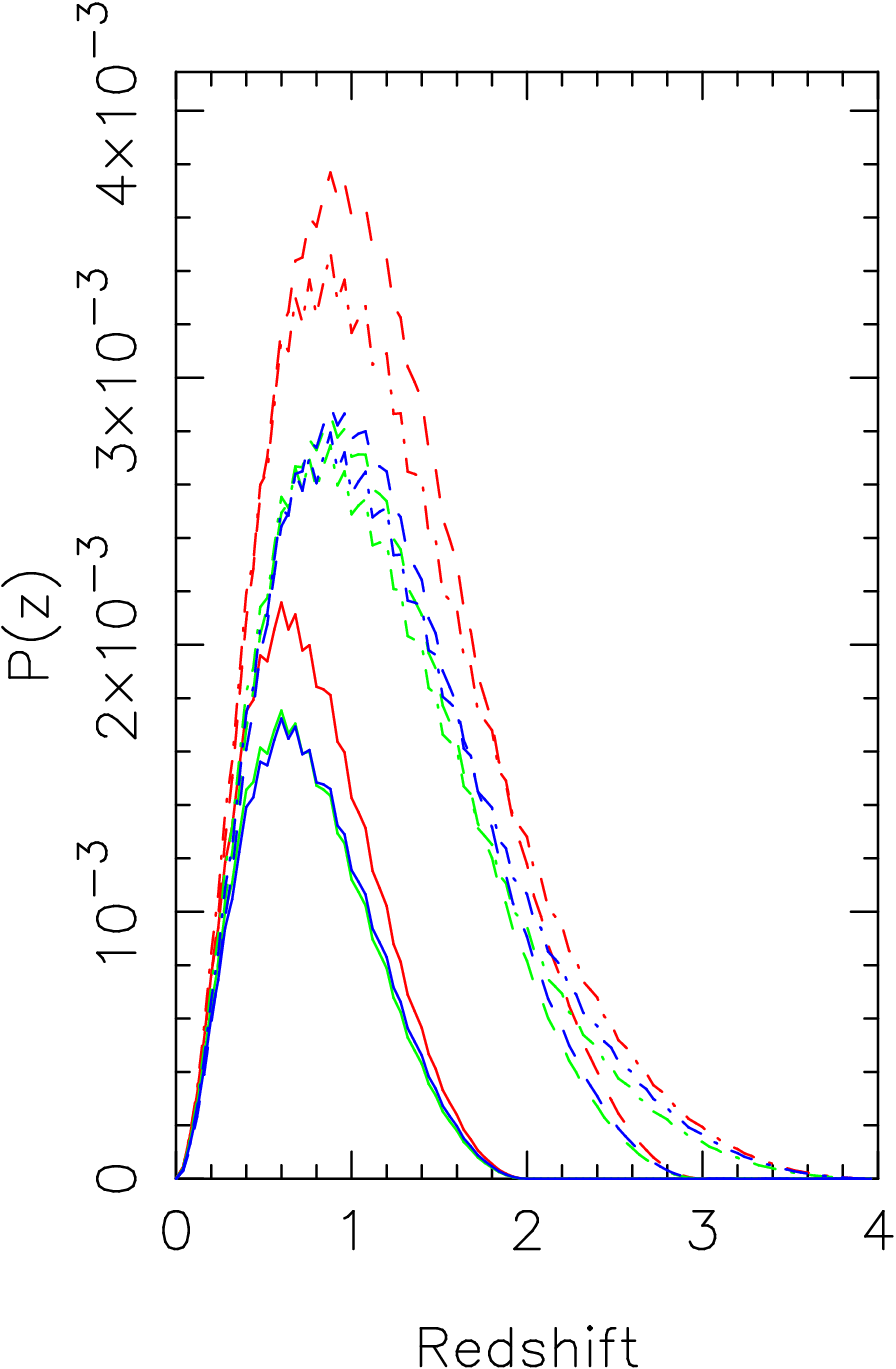}
  \caption{Plot of the the probability of strong lensing ($\mu > 2$)
as a function of lens redshift for three source redshifts: $z=2$ (solid lines),
$z=3$ (dashed lines), $z=4$ (dot-dashed lines).
The red lines are the predictions for the PS halo mass function,
the blue lines are those for the ST halo mass functions, and the
green lines those for the T halo mass functions.
The slight lumpiness to the lines is a numerical artefect.
}
\end{figure}

\begin{figure}
\includegraphics[width=84mm]{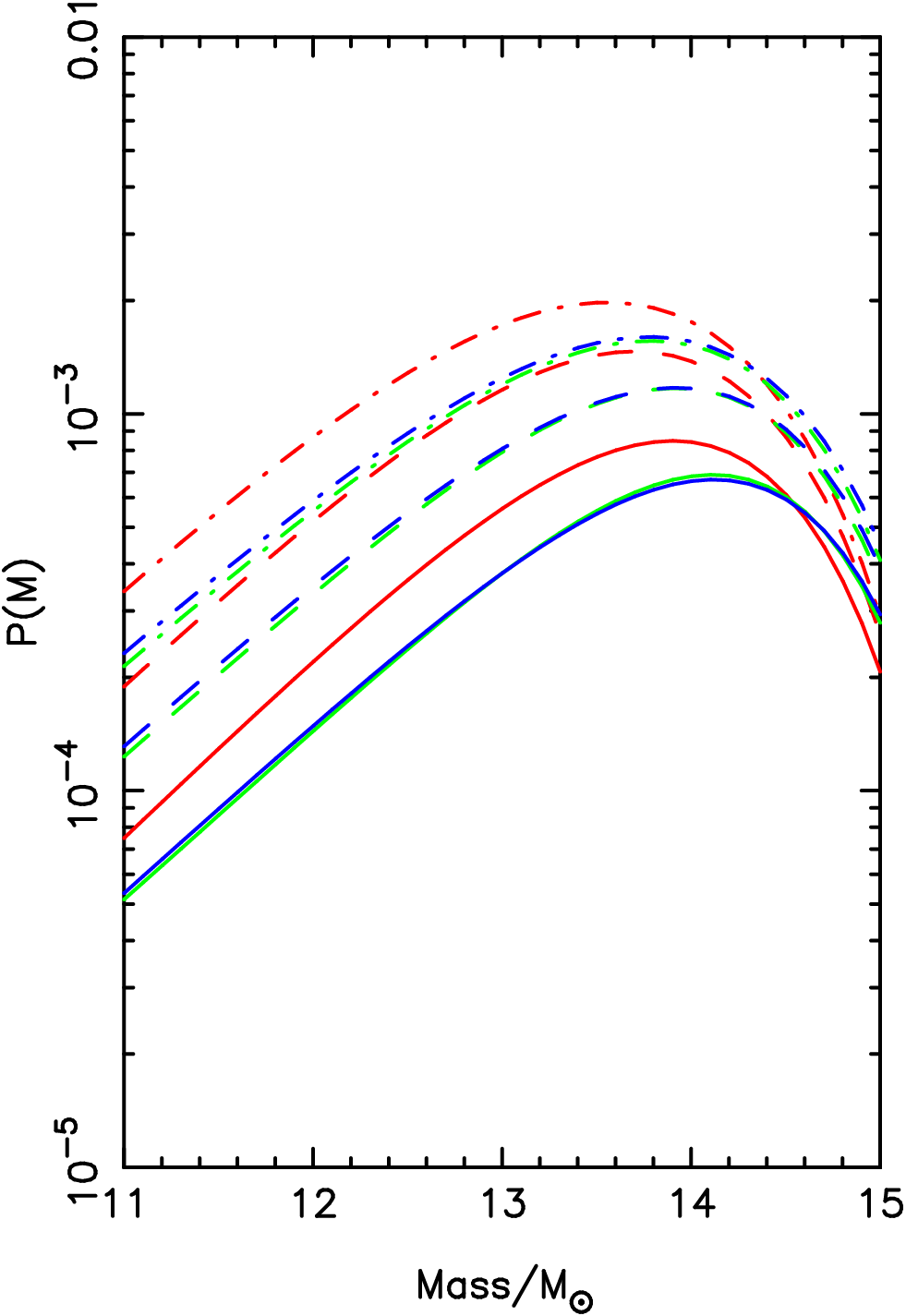}
  \caption{Plot of the probability of strong lensing ($\mu > 2$)
as a function of halo mass 
for three source redshifts: $z=2$ (solid lines),
$z=3$ (dashed lines), $z=4$ (dot-dashed lines).
The red lines are the predictions for the PS halo mass function,
the blue lines are those for the ST halo mass functions, and the
green lines those for the T halo mass functions.}
\end{figure}

I have made a first attempt to test the different halo mass functions using the results
from Bussmann et al. (2013). This paper describes observations of 30 Herschel sources
with $\rm S_{500 \mu m} \geq 100 mJy$ which are not either blazars or associated with
low-redshift star-forming galaxies. Models suggest that sources that obey these criteria
are highly likely to be lensed systems (Negrello et al. 2010) and so this sample
is a good test of whether
it is possible to use the properties of lensed sources to place useful constraints
on the halo mass function. 
I have removed two sources from the sample (Hermes J022016.5-060143 and H-ATLAS
J084933.4+021443) because there is evidence that the magnification is $<$2, the
threshold for strong lensing.
For the 28 remaining systems, there are 26 spectroscopic redshifts for the sources, 20
spectroscopic redshifts for the lenses and 23 measurements of the Einstein radius.
For two sources, there is evidence for more than one lens, showing that practical
cosmology experiments will need to take account of multiple lensing; in this
initial investigation I have simply averaged the measurements. 
Bussman et al. (2013) assumed that density profiles
of the lenses were singular isothermal ellipsoids, which is close enough to the
assumption made in this paper given the exploratory nature of the
present investigation.

Figure 5 shows the predicted distribution of $\sum_i p_{model}(z_L|z_{s,i})$ 
for the sample, where the sum is over the objects in the sample. 
The individual probabilities are given by

\smallskip
\begin{align}
p_{model}(z_L|z_{s,i}) = \int_2^{\infty} \int_{M_l}^{M_u} p(z_l,\mu,M|z_s) dM d\mu
\end{align}
\smallskip
\noindent where the probability on the right-hand side is given by equation 3.
Figure 5 also show the
histogram
of the measured lens redshifts, which should reflect, if the model is correct,
the probability distribution shown in the figure.
Figure 6 shows the predicted distribution
of $\sum_i p(\theta_E|z_{s,i})$, where the probabilties are calculated
in a similar way to equation 12, and the histogram of the measured Einstein radii.
The predicted distributions are very similar for the different halo mass
functions. In both figures, there appears to be a discrepancy between the
predictions and the observations.

To determine whether the differences are statistically
significant, I have invented the following statistics:

\smallskip
\begin{align}
S_1 = \int_0^{z_{L,i}} p_{model}(z_L | z_{S,i}) dz_L 
\end{align}
\smallskip

\noindent and

\smallskip
\begin{align}
S_2 = \int_0^{\theta_{E,i}} p_{model}(\theta_E | z_{S,i}) d\theta_E 
\end{align}
\smallskip

\noindent in which $z_{L,i}$, $z_{S,i}$ and $\theta_{E,i}$ are the observed 
lens redshift, source
redshfit and Einstein radius of the i'th source.
The ensemble of values of $S_1$ and $S_2$ for the sample are empirical
statistics describing how well the data are described by
the underlying model. For example, consider the redshifts of the lenses
(equation 13). If the redshifts of the lenses are systematically
lower than the predictions of the model, the values of $S_1$ will tend
to fall below 0.5; if the observed redshifts are systematically higher than the
model, they will tend to be above 0.5.
If, on the other hand, the model represents the data well, the values of
$S_1$ and $S_2$ should be
uniformly distributed between 0 and 1.
This is, of course, based on the assumption that there is no obvious selection
effect. An obvious possibility is that we might only recognise that a source
is lensed if the apparenent magnitude of the lens is brighter than some limit,
for example the magnitude limit of the Sloan Digital Sky Survey. In this case,
there would be a clear bias and the values of $S_1$ would be lower than they
should be.

I have tested the null hypothesis that the model represents the data by
applying a Kolmogorov-Smirnov one-sample test, in which 
the measured values of $S_1$ and $S_2$ are compared against the expected uniform distribution.
For the lens redshifts, the 
discrepancy between the values of $S_1$ and the expected uniform
distribution is
not formally significant (P$>$10\%)
for all halo mass functions. For the Einstein radii, the discrepancy between the values
of $S_2$ and the expected uniform distribution 
is significant at the $\simeq$1\% level (Kolmogorov-Smirnov
test).
The implications of these results are discussed in Section 5.

\begin{figure}
\includegraphics[width=84mm]{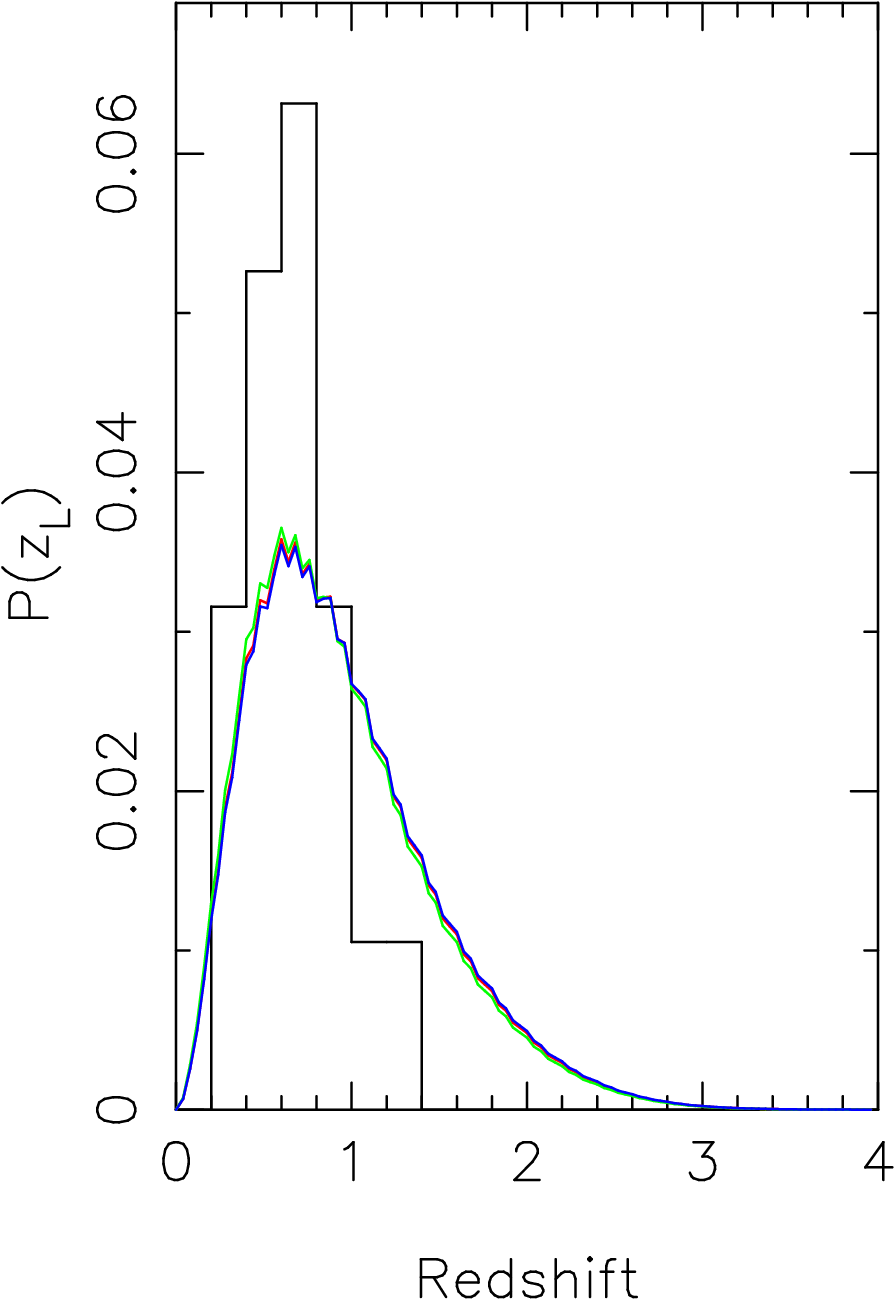}
  \caption{Plot of the predicted distribution of $\sum_i p_{model}(z_L|z_{s,i})$
using the method of Section 2.1 for the sample of 
lensed Herschel sources of Bussmann et al. (2013).
The green line is the prediction for the
T halo mass function and the blue line for the ST halo mass function.
The prediction for the PS halo mass function is virtually identical
to the prediction for the ST mass function.
The histogram shows the measured lens redshfits.}
\end{figure}

\begin{figure}
\includegraphics[width=84mm]{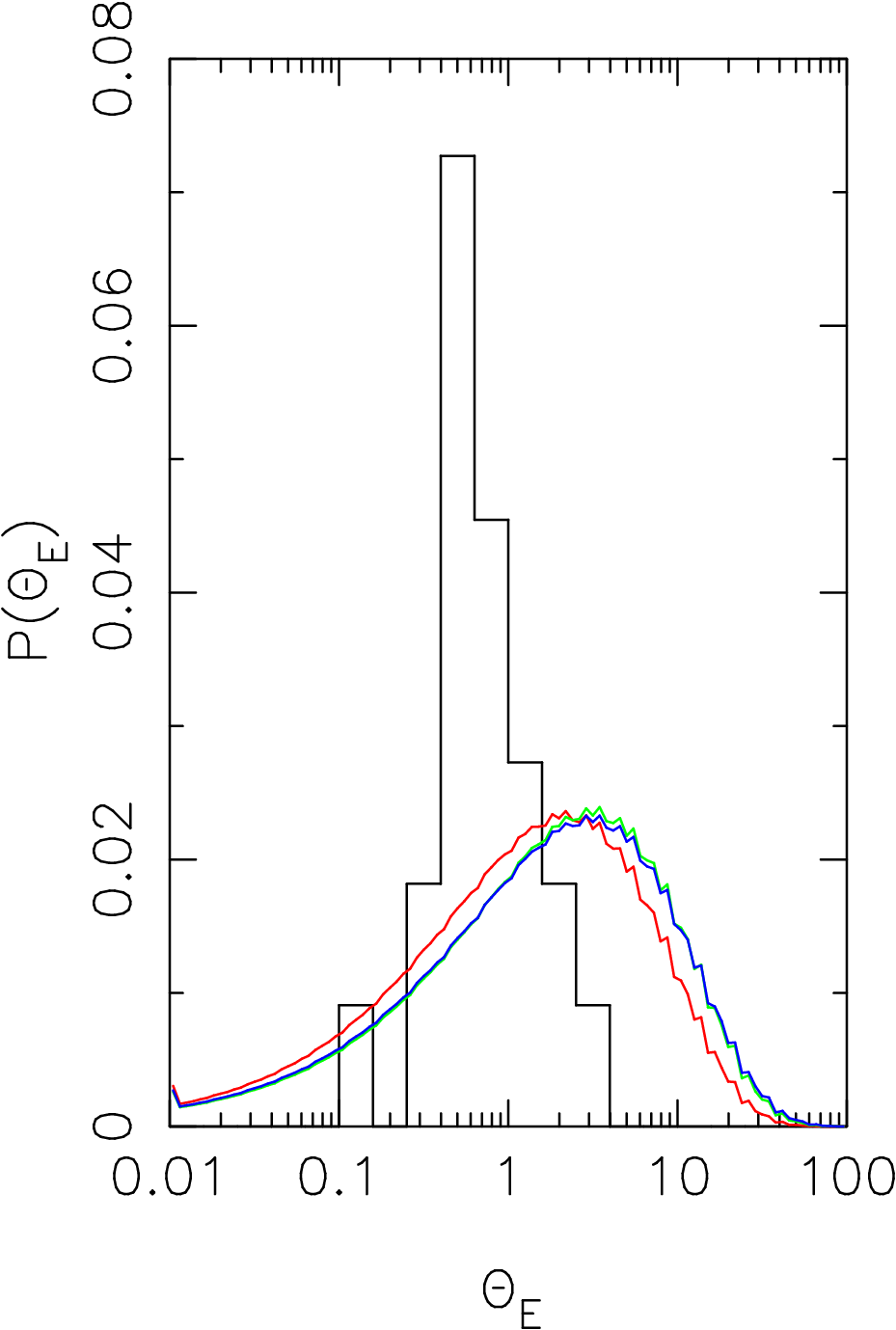}
  \caption{Plot of the predicted distribution of $\sum_i p_{model}(\theta_E|z_{s,i})$
using the method of Section 2.1 for the sample of lensed Herschel sources of Bussmann et al. (2013).
The red line is the prediction for
the PS halo mass function, the green line for the
T halo mass function and the blue line for the ST halo mass function.
The histogram shows the measured Einstein radii.}
\end{figure}

\subsection{A universal density profile}

The previous method used the velocity dispersion of the halo as a stepping stone
from the mass of the halo to the properties of the gravitational lens. Since
the physical quantity that the lensing measures most
directly is the mass interior to the Einstein radius, a more direct technique
to connect the mass and lensing properties of a halo is
to make some assumption about the density profile of the halo.
A complication has been that lensing studies show that the density profiles of
lenses generally follow an SIS density profile, while N-body simulations imply that the
dark matter should follow an NFW profile (Section 1). Recently, however, Lapi et al.
(2012) have shown that this kind of behaviour is expected from the combination
of baryons and dark matter that should exist in real halos; in the centre of a
halo the baryons dominate, and one naturally obtains an SIS density profile,
whereas the density profile approaches the NFW form in the outer part of the halo
where the dark matter dominates. They have proposed that a universal profile could
be used for lensing studies, and this is the approach I will use in this
section.

The universal profile depends on the mass of the halo, the ratio of the mass of
baryons to the mass of dark matter, the redshift of the halo, the Sersic index of the
stars, and the redshift at which the halo reached virial equilibrium.
I assume the parameters proposed by Lapi et al. (2012), with the exception of
the virialisation redshift, which I place at $z=4$ rather than $z=2.5$ in order
to move it above the redshifts of all the sources for the sample of Bussmann et
al. (2013). One complication is that the definition of the halo mass used
by Lapi et al. (2012), which is the mass out to the virial radius, is not
quite the same as the definition used in N-body simulations. Tinker et al. (2008),
for example, select halos by finding surfaces that obey the criterion:

\smallskip
\begin{align}
\Delta = {3 M(<R) \over 4 \pi R^3 \rho_b}, 
\end{align}
\smallskip
\noindent where $\rho_b$ is the background density and $\Delta$ is the
chosen over-density. However, given a universal profile, it
is easy to calculate the difference between this mass and the mass within
the halo virial radius used by Lapi et al. (2012).
I only consider the
halo mass function ($\Delta=200$) from Tinker et al. (2008), since this is the
only one where the information exists to make a quantitative 
extrapolation of this sort.

From the density profile for a halo at a given redshift and mass, it
is straightforward to calculate the cross-section for strong lensing
and the Einstein radius from the critical surface density for
strong lensing, $\Sigma_c$:

\smallskip
\begin{align}
{M(<r_c) \over \pi r_c^2} = \Sigma_c = {c^2 D_s \over 4 \pi G D_{LS} D_S} 
\end{align}
\smallskip
\noindent where $M(<r_c)$ is the mass interior to a projected radius $r_c$ on the sky and 
\smallskip
\begin{align}
\theta_E = {r_c \over D_L} 
\end{align}
\smallskip
\noindent Figure 7 shows the predicted distribution of $\sum_i p_{model}(z_L|z_{s,i})$
for the sample from Bussmann et al. (2013), where the sum 
is over the objects in the sample, together with a histogram
of the measured lens redshifts. Figure 8 shows the predicted distribution
of $\sum_i p_{model}(\theta_E|z_{s,i})$ and the histogram of the measured Einstein radii. 
Using the same statistics as the previous section, 
for the redshifts the measured values of $S_1$ are not
significantly different from the predicted
uniform distributon ($>$10\%, Kolmogorov-Smirnov one-sample test). In contrast,
for the
Einstein radii the measured values of $S_2$ are significantly different
from the predicted uniform distribution
($<$1\%, Kolmogorov-Smirnov test).
We will discuss these results and the results of Section 2.1 in Section 5.

\begin{figure}
\includegraphics[width=84mm]{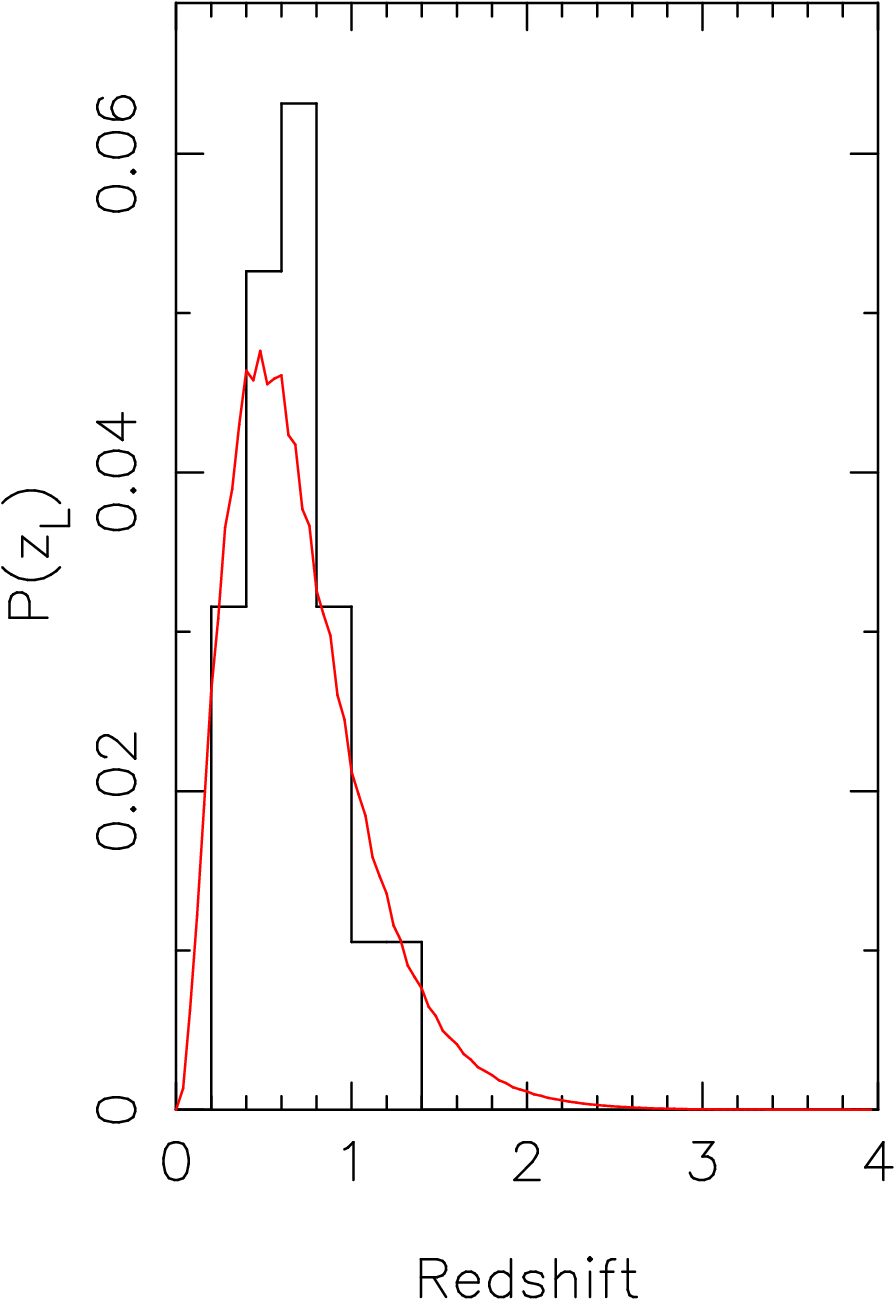}
  \caption{Plot of the distribution of $\sum_i p_{model}(z_L|z_{s,i})$
using the method of Section 2.2 for the sample of lensed Herschel sources of Bussmann et al. (2013),
predicted using the universal profile of Lapi et al. (2012) and the
halo mass function from Tinker et al. (2008).
The histogram shows the measured lens redshfits.}
\end{figure}

\begin{figure}
\includegraphics[width=84mm]{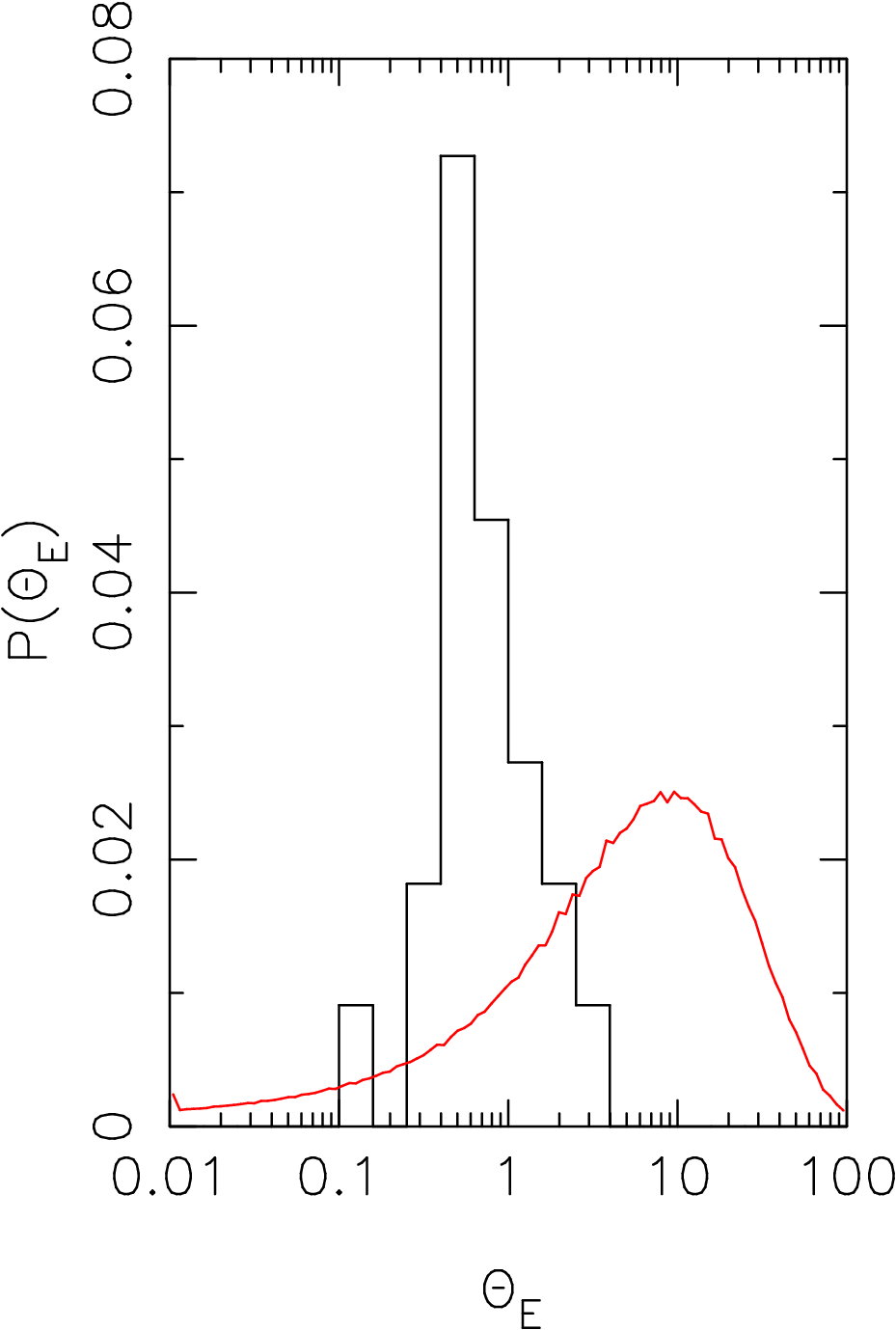}
  \caption{Plot of the distribution of $\sum_i p_{model}(\theta_E|z_{s,i})$
using the method of Section 2.2
for the sample of lensed Herschel sources of Bussmann et al. (2013),
predicted using the universal profile of Lapi et al. (2012) and the halo mass
function of Tinker et al. (2008).
The histogram shows the measured Einstein radii.}
\end{figure}

\section[]{Cosmological Parameters}
 
Grillo et al. (2008) have suggested a way of estimating cosmological parameters
that is independent of the distribution of dark-matter halos and the properties
of the sources. In this section, I investigate whether this method is practical,
given the sensitivities of current telescopes, for estimating 
$\Omega_{\Lambda}$ and $w$, the parameter in the equation-of-state of dark energy.

In the method proposed by Grillo et al., the four quantities that need to be
measured are the redshifts of the lens and source, the Einstein radius and
the velocity dispersion of the lens. They assume, as I do here, that the density
profile of the lens is a singular isothermal sphere, and show that
the central velocity dispersion of the stellar component of the lens
is a good measure of the SIS velocity disperson. In this case

\smallskip
\begin{align}
r(z_L,z_S; \Omega_{\Lambda}, \Omega_M) = {D_{LS} \over D_{S}} = {c^2 \theta_E \over
4 \pi \sigma_0^2}, 
\end{align}
\smallskip
\noindent in which $\sigma_0$ is the central stellar velocity dispersion. Given
measurements of $\sigma_0$, $z_L$, $z_S$ and $\theta_E$, we can then use this
relationship to estimate $\Omega_M$ and $\Omega_{\Lambda}$. 

To test the method I have carried out a Monte-Carlo simulation, generating artificial
samples of sources,
in which 
the source redshifts, lens redshifts and Einstein
radii are drawn from uniform probability distributions
over the ranges
$2 < z_s < 4$,
$0.4 < z_L < 1.4$ and $1.0 < \theta_E < 1.5$ arcsec.
These ranges reflect fairly well the ranges found for real Herschel lensed
systems (Bussmann et al. 2013).
From each triplet of Einstein radius, lens redshift and source redshift generated from
the Monte-Carlo simulation,
I have calculated the velocity dispersion of the lens from
the relationship between velocity dispersion and Einstein radius for
an SIS lens:

\smallskip
\begin{align}
\sigma^2 = {c^2 D_S \theta_E \over 4 \pi D_{LS}} 
\end{align}
\smallskip

\noindent As the background cosmology, I assume the Planck cosmology:
a flat universe ($\Omega_{\Lambda} + \Omega_M = 1$) and a value of $\Omega_M$ of
0.315. 
I then simulate the observations of each lensed system, using realistic estimates
of the accuracy with which the observations can be made.
I assume that the lens and source redshifts are measured precisely
but that the Einstein radius is measured with a precision of 5\% (Dye et al. 2014)
and that the stellar velocity dispersion is measured with a precision of
5\%, similar to the accuracy estimated by Treu and Koopmans (2004)
for their measurements of the
stellar velocity dispersions of gravitational lenses. 
Using this method, I generate 100 artificial samples, each containing 100
sources.

Using the simulated
observations, I then estimate the cosmological parameters from
each lens sample, 
minimising the chi-squared discrepancy between the second and third terms in
equation 18. Equation 3
of Grillo et al. (2008) gives the full expression for chi-squared. I assume that
the Universe is flat, so that $\Omega_M + \Omega_{\Lambda} = 1$ and
so there is one cosmological parameter to estimate.
Figure 9 shows the estimates of $\Omega_{\Lambda}$ for a Monte-Carlo simulation of
100 samples, each containing 100 lensed systems. 
The mean value of $\Omega_M$ for the samples is 0.37 with a standard deviation,
calculated from the mean of the estimates for the 100 samples rather than
from the true cosmological value,
of 0.03. Thus the statistical precision of an estimate of
$\Omega_M$ using this method, even from a sample of only 100
lensed systems, is high, although there is a systematic
shift from the true value of $\Omega_M$, a phenomenon also seen in the
simulations of Grillo et al. (2008).

The probable reason for the systematic shift is 
the $\sigma_0^2$ term in equation 18 because any error in the velocity
dispersion, whether positive or negative, will produce a shift in $\Omega_M$ in the same
direction. To check that the results are not sensitive to the assumptions about the 
uniform probability distributions assumed in the Monte-Carlo simulation
I carried out a second Monte-Carlo simulation, this time
using a 2D conditional probability distribution
for the lens redshift and Einstein radius given the source redshift, $p(z_L,\theta_E|z_s)$,
derived from the halo mass function of Tinker et al. (2008) and the method of \S 2.1.
I used a fixed source redshift of 3. This time the mean value of the estimates
of $\Omega_M$ for the 100 samples was 0.36 with a standard deviation of 0.03, very similar
to the results from the first simulation.

It is
easy to extend this method to more complex cosmologies. Suppose that the universe
is flat but that $w$, the index in the equation-of-state of dark energy, is not
-1, the value for a cosmological constant. In this case, there are two
cosmological parameters, $w$ and $\Omega_M$, to estimate. As the background
cosmology, I assume a value of $\Omega_M$ of 0.315 and 
a value for $w$ of
-1.5, similar to the value obtained from Planck observations alone (Planck Collaboration 2013). 
To generate a lensing probability model using this cosmology, I have started with the
equations given by Weinberg et al. (2013):

\begin{align}
{H^2(z) \over H_0^2} = & \Omega_M(1+z)^3 + & \Omega_r(1+z)^4 \nonumber \\
&  + \Omega_k(1+z)^2 + & \Omega_{\phi} {u_{\phi}(z) \over u_{\phi}(z=0) }
\end{align}

\begin{align}
{u_{\phi}(z) \over u_{\phi}(z=0) } = (1+z)^{3(1+w)}.
\end{align}

\noindent I assume $\Omega_k$ and $\Omega_r$ are both zero,
and so, from equation 19, $\Omega_M + \Omega_{\phi} = 1$. The
fundamental equation, equation 3, then becomes:

\smallskip
\begin{align}
 p(z_l, \mu, M | z_s) dz dM d\mu = &  \nonumber\\
  {c n(z_l,M)(1+z_l)^2 \over
  H_0 \sqrt{(1+z_l)^3 \Omega_M + (1-\Omega_M)(1+z_l)^{3(1+w)} } } & {d\sigma(>\mu) \over d\mu} & \nonumber\\
   dz dM d\mu, &  
\end{align}
\smallskip

Figure 10 shows the estimates of $w$ obtained from a Monte-Carlo simulation
of 100 samples, each containing 1000 lensed systems. 
The Monte-Carlo simulation used the same uniform probability distributions as above.
The mean value of the estimates of $w$ from the 100 samples is -1.40 with a standard deviation around
this mean of
0.37.
Therefore, even samples containing 1000 lensed systems would not be enough
to estimate $w$ with high accuracy, although the accuracy is at least as good as
was obtained from the Planck observations alone (Planck Collaboration 2013), showing
that for investigating dark energy large lens samples are competitive with
observations of the cosmic microwave background.

\begin{figure}
\includegraphics[width=84mm]{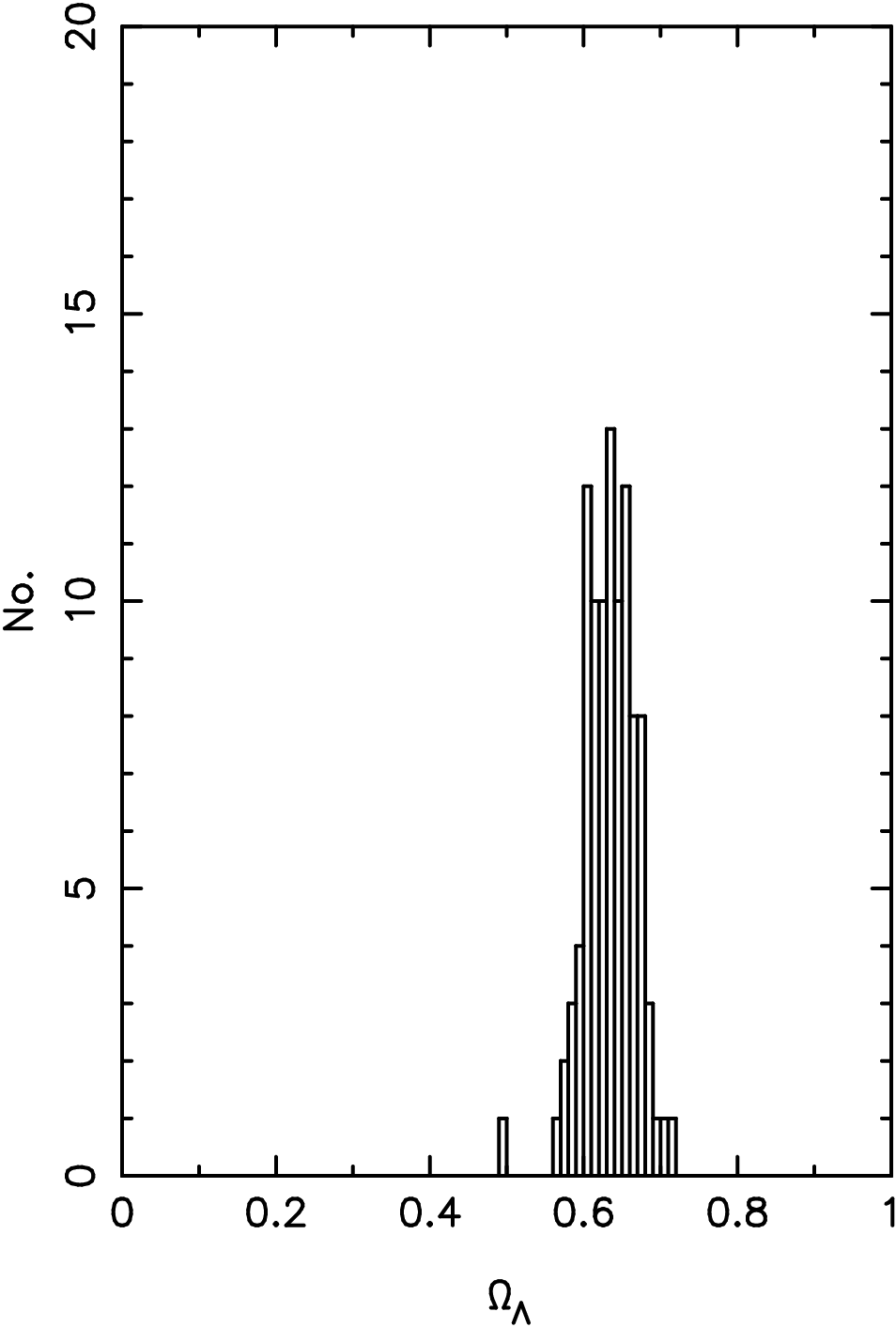}
  \caption{Plot of the values of $\Omega_{\Lambda}$ recovered
from a Monte-Carlo simulation of 100 samples, each containing 100
lensed systems, using the method described in the text.}
\end{figure}

\begin{figure}
\includegraphics[width=84mm]{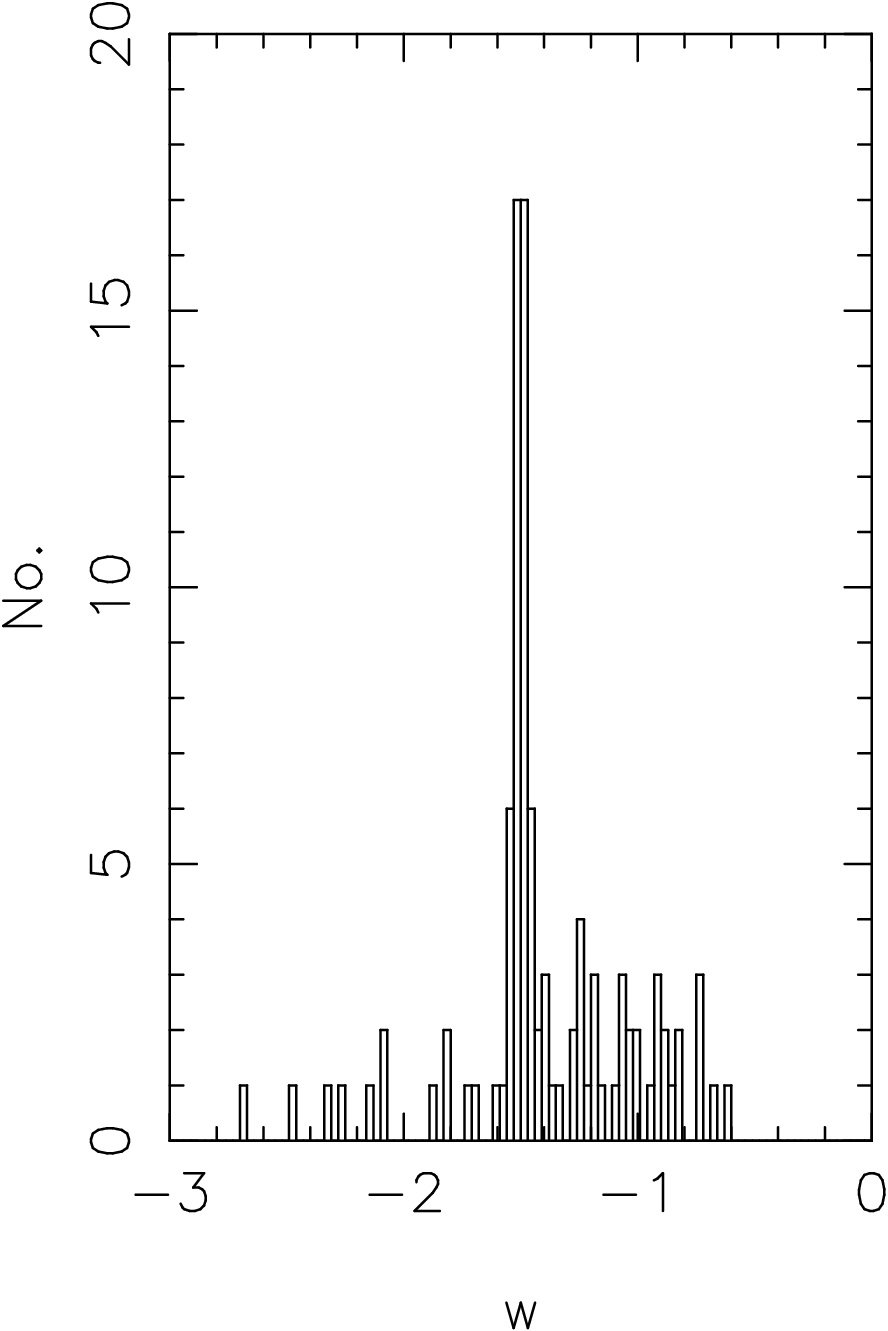}
  \caption{Plot of the values of $w$, the parameter in the equation-of-state
of dark energy, recovered from
a Monte-Carlo simulation of 100 samples, each containing 1000
lensed systems, using the method described in the text.}
\end{figure}

\section{The Population of Sources}

In this section I show that there is one property of the source population that
surprisingly is largely independent of the statistical properties of the
dark-matter halos. This is the fraction of sources at a given submillimetre
flux and redshift that have been strongly lensed. 
Also rather surprisingly, 
the thing that this property does most strongly depend on is the amount of evolution
that the sources themselves have undergone.
The fraction of lensed
sources is something
that in principle can be measured, since any source that is strongly
lensed (magnification $>$2)
will have at least two images.

The models of Negrello et al. (2007, 2010) suggest
that the lensing fraction is only high at very bright submillimetre flux
densities. However, this model is a single very specific model, albeit one that gives
good agreement with the submillimetre source counts. In this section, I present a more
general analysis to determine how this fraction depends on different assumptions about
the halo mass function and galaxy evolution.

As before, I assume that the lenses have an SIS density profile and the
standard Planck cosmology. In this case, the
probability that a source at redshift $z_s$ is magnified by a factor $\mu$ is given by

\smallskip
\begin{align}
p(\mu,z_s) = {p_0(z_s) \over \mu^3}, 
\end{align}
\smallskip

\noindent I obtain the numerator on the right-hand side by integrating this equation
over magnification and equation 3 over mass, lens redshift and magnification, as
follows:

\smallskip
\begin{align}
\int_0^{z_s} \int_2^{\infty} \int_{M_l}^{M_h} p(z_l,\mu,M|z_s) & dM d\mu dz_l \nonumber\\ 
= \int_2^{\infty} {p_0(z_s) \over \mu^3} & d\mu
\end{align}

\smallskip 
\noindent I calculate $p_0(z_s)$ for each of the three
halo mass functions
in Section 2. 

For a galaxy at a given redshift, $z_s$, the probability that the galaxy is lensed decreases strongly
with magnification (equation 23). However, in the real Universe we can't observe all the
galaxies at a redshift $z_s$. Instead, we might observe all the galaxies with a flux density greater
than a flux density $S_{lim}$ at this redshift. A thought experiment shows that we might find that a very
high fraction of the galaxies are strongly lensed. Suppose $L_{lim}$ is the luminosity corresponding
to the redshift of interest, $z_s$, and the flux density limit, $S_{lim}$. Suppose that
there are no galaxies in the Universe with intrinsic (unlensed) luminosities above this limit.
The only galaxies that we will find at this redshift above this flux density limit will then
be lensed
sources. This is why the method of Negrello et al. (2010) is so efficient at finding lensed
systems because Negrello et al. use a wavelength (500$\mu$m) and a flux density limit above which
there are virtually no high-redshift galaxies that will be detected unless they are strongly
lensed. Of course, this idea of a threshold luminosity is unrealistic, but the standard
galaxy luminosity function does drop off very rapidly at high luminosities. We will
now try and construct a detailed empirical model of this effect.

Before we consider the luminosity function, we will consider the
fraction of lensed sources that will be found as a function of observed flux density.
Let us suppose that $dN/dS(S_{obs})$ are the unlensed differential source counts at some
particular flux density $S_{obs}$. The fraction of sources,
$f(\mu)$, that will be found
at this flux density that will actually have lensing magnification $\mu$ is approximately
given
by:

\smallskip
\begin{align}
f(\mu) = p(\mu,S_{obs}/\mu) { {dN \over dS} \left( {S_{obs} \over \mu} \right) \over {dN \over dS} \left( S_{obs} \right) } \times
{1 \over \mu^2}
\end{align}
\smallskip

\noindent In this equation, the first term is the probability that a source
of flux density, $S_{obs}/\mu$ will be magnified by a factor of $\mu$.
In equation 25, 
one factor of $\mu$ arises because gravitational lensing increases
the solid angle of the source plane, thus reducing the surface density of sources. The other
factor of $\mu$ arises because the differential $dS$ is being magnified by a factor $\mu$.
However, despite these two factors of $\mu$ that decrease the fraction of highly magnified
sources, and the low probability that an individual source is
lensed, it is often possible to find highly magnified sources because the differential
source counts increase rapidly with decreasing flux density. 

We assume that at any redshift, the luminosity function of the unlensed
sources is given
by a standard Schechter function:

\smallskip
\begin{align}
\phi(L) = {\phi_* \over L_*} \left( {L \over L_*} \right)^{\alpha_S} exp\left({ -{L \over L_*}}\right)
\end{align}

\smallskip

\noindent In this equation, $L_*$ and $\phi_*$ may change with redshift, although in
practice it is only the evolution of $L_*$ that is important.
We will see that the fraction of sources at some redshift above a chosen flux density
depends critically on the evolution of $L_*$, and that therefore the fraction of
lensed sources actually gives us very useful information about the strength of the
cosmic evolution.

Equation 25 describes the effect of magnification on the source counts no matter
what the redshifts of the sources. We will now consider the source counts at a
particular redshift.
For a fixed redshift, the observed flux density of a source is proportional
to its luminosity. Thus the relationship between $dN/dS$ 
and flux density has the same form as the relationship between the luminosity function
and luminosity. Let us define $L_{obs}$ as the luminosity corresponding to
the chosen flux density, $S_{obs}$ at the chosen redshift, $z_s$. We can then write
the fraction of sources at this redshift with this flux density, $S_{obs}$, which
are magnified by a factor $\mu$ as

\smallskip
\begin{align}
f(z_s,S_{obs},\mu) = p(\mu,z_s) {\phi \left( {L_{obs} \over \mu}\right) \over \phi (L_{obs}) } \times {1 \over \mu^2}
\end{align}
\smallskip

\noindent The probability in the equation is the same as the probability we calculated in
equation 23. 
The probability no longer depends on flux density, as it did in equation 25, because all
sources at the same redshift are magnified by the same factor no matter what their
flux density.

The fraction of the sources at this redshift and flux density that are
strongly lensed is thus:

\smallskip
\begin{align}
f_{SL}(z_s,S_{obs}) = \int_2^{\mu_{max}} f(z_s,S_{obs},\mu) d\mu
\end{align}
\smallskip

\noindent Combining equations 23, 26 and 27, we obtain:

\smallskip
\begin{align}
f_{SL}(z_s,S_{obs}) = \int^{\mu_{max}}_2 p_0(z_s) \mu^{-(\alpha_S+5)} exp\left( {L_{obs}(\mu-1) \over \mu L_*} \right) d\mu 
\end{align}

\smallskip

\noindent We use this expression to calculate the fraction of lensed
sources as a function of redshift and flux density at 250$\mu$m, the main wavelength
of many of the Herschel extragalactic surveys.
I have mapped $L_{obs}$ to flux $S_{obs}$ at a given redshift using the spectral energy distribution derived by
Pearson et al. (2013) for high-redshift Herschel sources.

Our knowlege of the evolution of the submillimetre luminosity function is still relatively
poor (Eales et al. 2010; Gruppioni et al. 2013). 
In particular, we know little about whether there is any evolution of the
low-luminosity slope of the luminosity function (Gruppioni et al. 2013), so I
have made the simplifying assumption that $\alpha_S$ is constant with
redshift.
I have used two different models for the evolution. Gruppioni et al. (2013)
have presented empirical luminosity functions at a rest-frame wavelength of 
90$\mu$m in redshift bins up to a maximum redshift of 4.5. I have fitted a Schechter
function to the empirical luminosity function in each bin. For their lowest redshift bin,
I allowed $\alpha_S$ to vary and found a good fit with $\alpha_S=-1.3$; in the bins at higher
redshift, I have kept $\alpha_S$ fixed with this value. Using this method, I have derived
a value for $L_*$ in each bin. The dependence on redshift of $L_*$ is represented well
by $L_* \propto (1+z)^{2.89}$ out to $z=3.5$, with $L_*$ remaining constant
at higher redshifts. This is our first model for the evolution of the luminosity
function. Our second model comes from the work of Eales et al. (2010),
who found strong evolution in the 250-$\mu$m luminosity function over the
redshift range $0 < z < 1$ but no evidence for any evolution at higher redshift. 
I have fitted a Schechter function to the luminosity functions in each bin in the same way as
for the other dataset, which resulted in a model with
$L_* \propto (1+z)^{4.15}$ out to $z=1$, with no evolution at higher redshifts.

The lines in Fig. 11 show the position on the flux-redshift plane where 
the models predict that different fractions of sources should be
lensed. For the coloured lines, the different line styles correspond
to four different halo mass functions, the three used in Section 2
and an extreme one
in which I have taken the halo mass function of Tinker et al. (2008)
and multiplied the values in every mass bin by a factor of 2.
The blue and red/green lines correspond to the predictions for
the two evolutionary models.
There is almost no difference for the different halo mass functions
and I also found negligible differences 
for different assumptions about
the maximum magnification.
The solid red and green lines show the predictions for a single model
for lensing fractions of 10, 30, 50, 70 and 90\%. These lines
show that the lensing fraction changes very rapidly with increasing
flux density.
The striking result in the figure is that it is the chosen evolutionary model
that makes a dramatic difference to the lensing fraction.
The physical reason for this
is shown by the other lines in the plot which show the position of $L_*$ for each
evolutionary model. At each redshift, the flux density at which the fraction of sources
that have been strongly lensed becomes significant corresponds to a 
luminosity $L \simeq 10 L_*$.  
Thus the fraction of lensed sources increases rapidly at the position on the diagram
where the luminosity function of unlensed sources is declining rapidly.

If the evolution found by Gruppioni et al.
(2013) is correct, it is only at the brighter flux densities that the fraction of
lensed sources is high, in agreement with the predictions of the model
of Negrello et al. (2007,2010), but if the weaker evolution found by Eales et al.
(2010) is correct, the fraction of sources that are lensed may be high at even quite faint
flux densities. This also suggests that an alternative way of investigating galaxy evolution
would be to measure the fraction of lensed sources as a function of flux and redshift.
Figure 12 shows the predicted distribution of magnification for sources with lensed luminosity
$L=10L_*$, 
$L=20L_*$ and $L=30L_*$.
The figure shows that most of the lensed sources at the flux threshold where the lensed fraction
is significant have quite modest magnifications, $\mu \simeq 3$.
For an SIS lens model, a source with this total magnification will have two
images with a flux ratio of 6. At brighter fluxes, the predicted magnifications are higher and
the flux ratios smaller. Observations with ALMA would be one
obvious way of measuring the fraction of sources that are lensed, as an alternative technique
of investigating galaxy evolution.

\begin{figure}
\includegraphics[width=84mm]{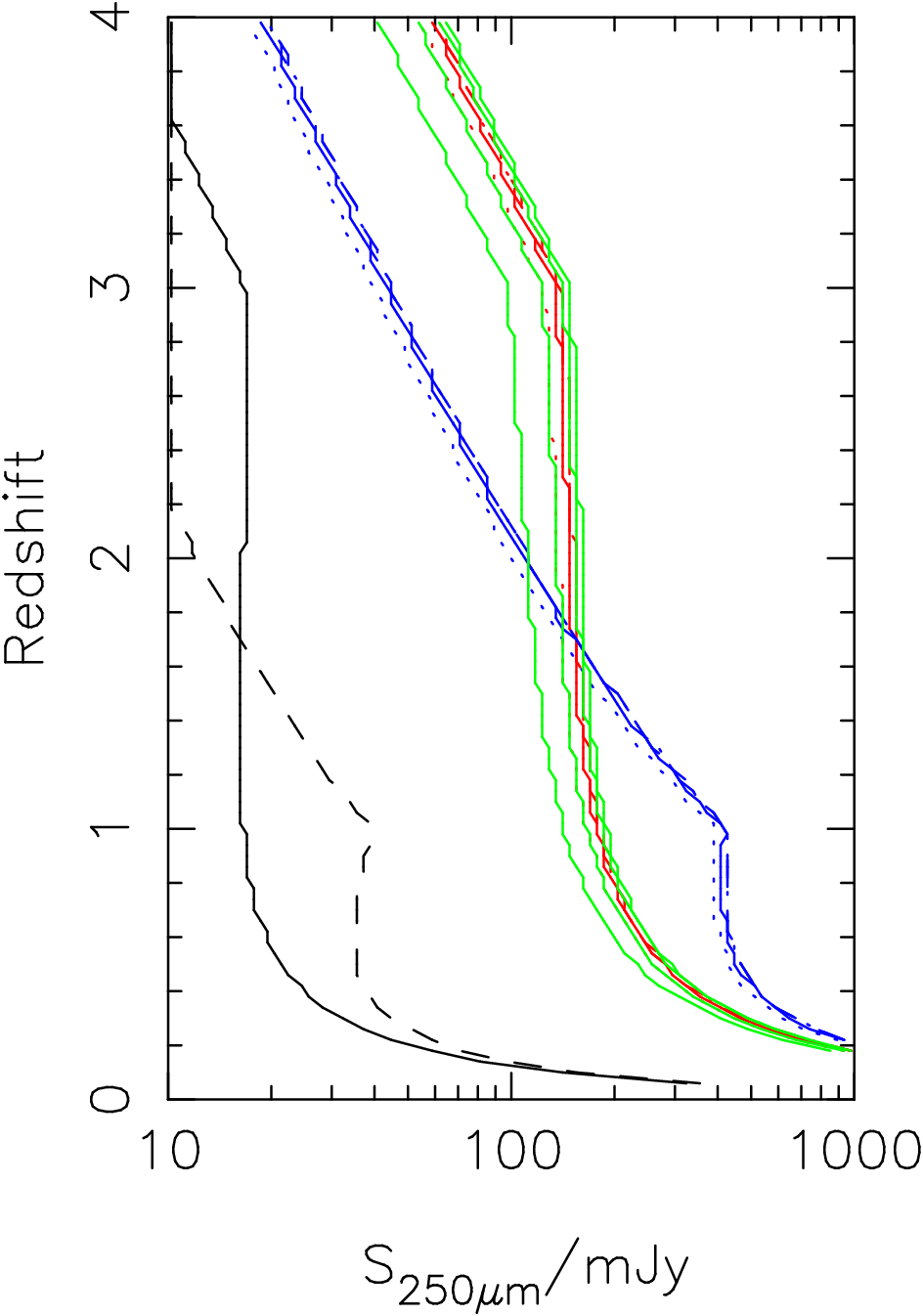}
  \caption{The coloured lines show the positions 
in the flux-redshift plane corresponding to different fractions
of lensed sources. The red and blue lines show the positions in which
50\% of the sources are lensed for eight different models.
The red lines show the predictions for a
model of galaxy evolution based on results
from Gruppioni et al. (2013) and the blue lines
a model based on the results of Eales et al. (2010) [See text for more details].
The line style corresponds to different halo mass functions: solid (Press and Schechter
1974), 
dashed (Tinker et al. 2008),
dot-dashed (Sheth and Tormen 1999), dotted (Tinker et al. multiplied by two, see
text). 
Note that the predictions for the Tinker et al. and Sheth and Tormen halo mass
functions are virtually the same.
The green lines show the predictions for the Gruppioni et al.
evolution model and the Press and Schechter halo mass function for lensing
fractions of 10\%, 30\%, 70\% and 90\%.
The solid and dashed dark lines show the position of an L* galaxy for the evolutionary models
based on the results of Gruppioni et al. and Eales et al., respectively.
}

\end{figure}

\section{Discussion}

My aim in this work has been to explore the best ways of using lensing samples to
investigate separately the values of cosmological parameters, the statistical properties
of the distribution of dark-matter halos and the properties of source population.
It is not possible to remove all degeneracies, but I have shown in Sections 2 and
4 that for an assumed cosmology it is possible to investigate separately the
statistical distribution of dark-matter halos and the properties of the source
population. Grillo et al. (2008) have shown that it is also possible to 
estimate cosmological parameters in a way that is independent of the source
population and the dark-matter halos. I have merely extended their work slightly,
showing that $w$, the parameter in the equation-of-state of dark energy, can also
be estimated using lens samples, although not very accurately.
I have also shown that the systematic errors in their method can be corrected in
a way that is independent of the distribution of dark-matter halos.
I have shown the fraction of sources that are lensed depends very weakly
on the properties of the intervening halos but very sensitivily on the
evolution of the source population. Therefore, measuring the fraction
of sources that are lensed as a function of flux density and redshift
may be a useful method of investigating galaxy evolution.

\begin{figure}
\includegraphics[width=84mm]{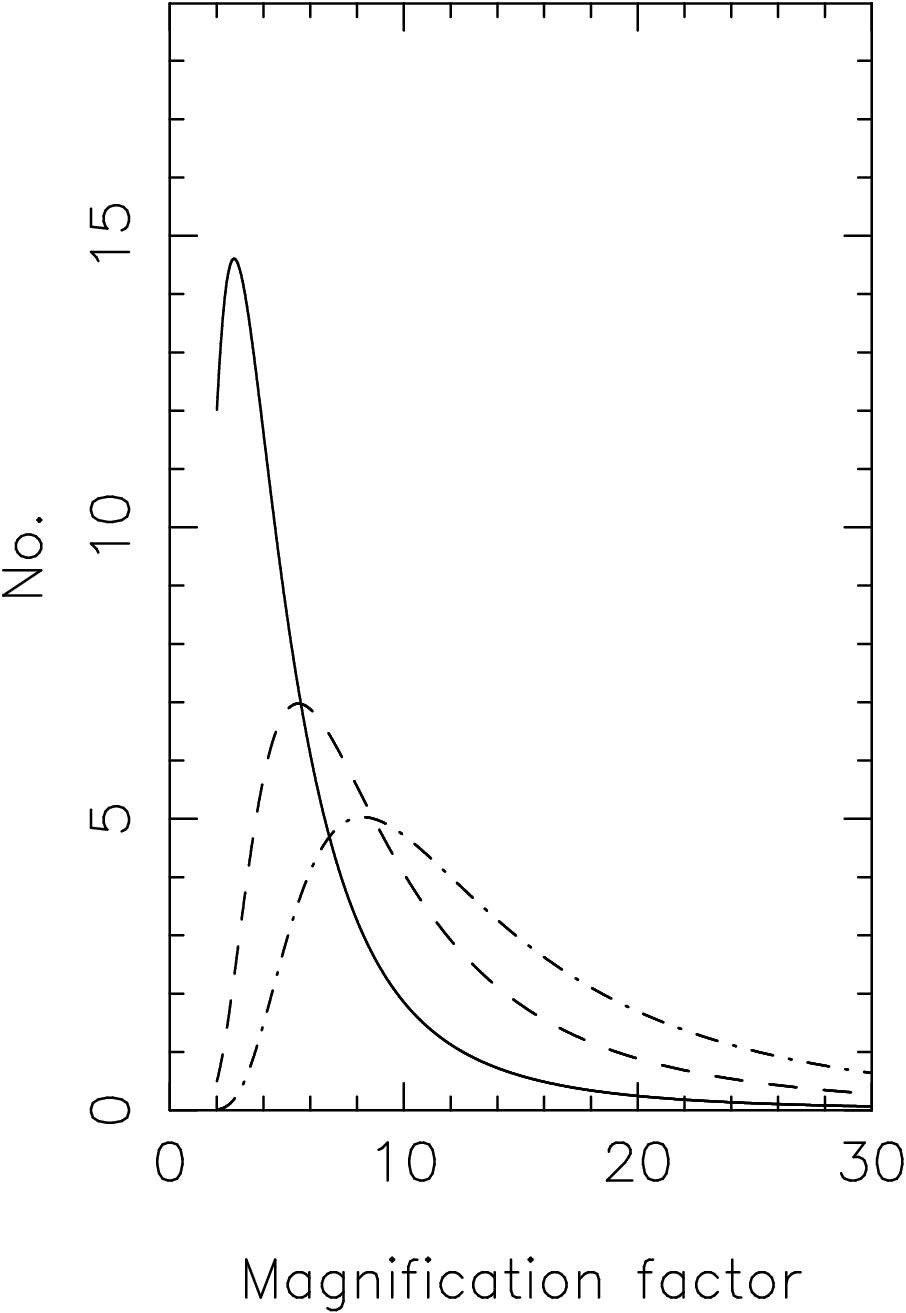}
  \caption{Plot of the predicted distribution of magnification factors for
strongly-lensed galaxies with different luminosities $L$ (after lensing).
The solid line shows the prediction for galaxies with $L=10L_*$, the dashed
line for galaxies with $L=20L_*$ and the dot-dashed line for galaxies
with $L=30L_*$.}

\end{figure}

We now return to the significant difference found in Sections 2.1 and 2.2 between
the models based on the standard statistical distributions of dark-matter halos
and the measurements of lensed Herschel sources presented in Bussmann et al.
(2013). It is possible that the relatively modest differences between the
measured and predicted lens redshifts (Figs 5 and 7) could be explained
by defects in the data. For example, not
all the lenses yet
have spectroscopic redshifts, and of course it is those at highest redshift
for which it is hardest to measure a redshift.
However, the difference in the measured and predicted Einstein radii
(Figs 6 and 8) is so large that it seems very unlikely that it could
be explained by either defects in the sample or the simplicity of
our assumptions, for example that all the lenses have an SIS density
profile.

However, this disagreement is not new. At the end of the previous generation
of lensing searches that used radio surveys it was realised that the
lensing statistics did not agree with the predictions of the halo models
(Keeton 1998;
Porciani and Madau 2000; Kochanek and White 2001). Rather than 
argue that the halo paradigm is wrong,
these authors explained the discrepancy as a combination of baryons not cooling in halos
above a critical mass and the perturbation of the
density distribution of the dark matter by the infalling baryons (Kochanek
and White 2001). Kochanek and White (2001) succeeded in replicating the
lensing statistics that then existed, although the baryon density to dark-matter density
required in the centres of halos, while in agreement with
the ratio measured in cosmological experiments, was higher than the
ratios measured for local galaxies.
We will leave for later attempts to reproduce the lensing statistics using detailed
models that incorporate baryonic physics. Here we point out that
the statistics for Herschel lenses is already much better than existed at the
end of the radio lensing era, and will soon get much better because of the
rapidly increasing number of lenses and because ALMA makes possible much
better observations of the lensed sources. Therefore, there is now a real
opportunity in this kind of study to carry out a detailed investigation
of the interaction of baryons and dark matter in the centre of halos.

\section*{Acknowledgments}

I thank  Asantha Cooray, Gianfranco De Zotti, Joaquin Gonzalez-Nuevo, Chris Kochanek,
Andrea
Lapi and
Mattia Negrello for useful suggestions and comments on the manuscript.
I am very grateful to Simon Dye for a meticulous reading of the manuscript.

\end{document}